\documentclass[a4paper,12pt]{article}
\usepackage[a4paper,width=150mm,top=25mm,bottom=25mm,bindingoffset=6mm]{geometry}
\usepackage[english]{babel}
\usepackage[pdftex]{graphicx}
\usepackage{placeins}
\usepackage{amsmath,amsfonts,amssymb,mathtools,bm,latexsym,stmaryrd}
\usepackage{rotating}
\usepackage{tabularx}
\usepackage[table,dvipsnames]{xcolor}
\usepackage{multicol,multirow}
\usepackage{relsize}
\usepackage[center]{subfigure}
\usepackage{caption}
\usepackage{slashed}
\usepackage{lineno}
\usepackage{cite,mcite}
\usepackage[title]{appendix}
\usepackage{url}
\usepackage[bookmarks=false,colorlinks]{hyperref}
\hypersetup{urlcolor=NavyBlue,citecolor=NavyBlue}
\usepackage{braket}

\newcommand{\ketbra}[2]{\mathinner{|{#1}\rangle\langle{#2}|}}
\def\I{\mathcal{I}}
\def\Y{\mathcal{Y}}
\def\Xp{{X^\prime}}
\def\Yp{{Y^\prime}}
\def\sift{\text{sift}}
\def\ver{\text{ver}}
\def\sec{\text{sec}}
\def\pa{\text{pa}}
\def\ec{\text{ec}}
\def\ph{\text{ph}}

\def\exp{\text{exp}}
\def\erf{\text{erf}}
\def\bin{\text{bin}}
\def\dc{\text{dc}}
\def\opt{\text{opt}}
\def\rhobar{{\overline{\rho}}}
\def\alphabar{{\overline{\alpha}}}
\def\phibar{{\overline{\varphi}}}
\def\thetabar{{\overline{\theta}}}

\begin{document}

\title{\bf\Large Security of the decoy-state BB84 protocol with imperfect state preparation}
\author{\sc Aleksei Reutov, Andrey Tayduganov, \\ \sc Vladimir Mayboroda and Oleg Fat'yanov}
\date{} 
\maketitle
\thispagestyle{empty}

{\centering \sl\small \noindent
Laboratory of Quantum Information Technologies, \\ National University of Science and Technology MISIS, \\ Moscow 119049, Russia
\par}

\vskip4cm
\begin{abstract}
    The quantum key distribution (QKD) allows two remote users to share a common information-theoretic secure secret key. In order to guarantee the security of a practical QKD implementation, the physical system has to be fully characterized and all deviations from the ideal protocol due to various imperfections of realistic devices have to be taken into account in the security proof. In this report, we study the security of the efficient decoy-state BB84 QKD protocol in the presence of source flaws, caused by imperfect intensity and polarization modulation. We investigate 
    non-Poissonian photon-number statistics due to coherent-state intensity fluctuations and the basis-dependence of the source due to non-ideal polarization state preparation. The analysis is supported by experimental characterization of intensity and phase distributions.
\end{abstract}

\pagebreak


\section{Introduction}

The BB84 protocol \cite{BB84} of quantum key distribution (QKD) between two distant parties, Alice and Bob, is based on the preparation and measurement of qubits in two bases (e.g., the computational $Z$--basis and the Hadamard $X$--basis), rotated relative to each other on the Bloch sphere by an angle of $\pi/2$. Any practical realization of the protocol inevitably introduces misalignments in the quantum state preparation and in the alignment of the measurement bases compared to the ideal ones, thus opening loopholes for information leakage to a potential eavesdropper (Eve). In particular, the preparation misalignments lead to the basis-dependence of the source that can allow Eve to distinguish the bases and attack two basis states separately. In the earliest security proof by Mayers \cite{Mayers96}, the source is assumed to be perfect. The simplified proof by Koashi and Preskill \cite{Koashi03} allows the source to be uncharacterized but basis-independent. The proofs by Lo and Chau \cite{Lo99}, Shor and Preskill \cite{Shor00} and Koashi \cite{Koashi09} are all based on the \textit{phase error} correction and assume the basis choice to be unknown to Eve. The phase error is a hypothetical error in the equivalent virtual entanglement-based protocol if the qubits of the EPR pair were measured in the complementary basis. It is well known that for an ideal source the phase and bit error rates in two bases are equal in the asymptotic limit, that is $E_Z^\text{phase}=E_X^\text{bit}$. However, if the source is basis-dependent, one cannot simply use one basis information to estimate the other. Thus, in this approach, the correct phase error rate estimation, taking into account the finite-key-size effects, is crucial for security analysis of the protocol with state preparation misalignments.

Further analysis by Gottesman et al. (referred to as GLLP) \cite{GLLP} proves the BB84 security for practical implementation of the QKD protocol when Alice's and Bob's devices are flawed. However, the GLLP approach is too conservative, it assumes the worst-case scenario when Eve can enhance the state preparation flaws by exploiting the channel loss. For this reason, the so-called \textit{loss-tolerant} QKD protocol was proposed by Tamaki et al. \cite{Tamaki14}, demonstrated experimentally \cite{Xu15,Tang16} and later on developed in Refs.~\cite{Mizutani15,Mizutani19,Pereira19,Pereira20,Curras21,Pereira23,Curras23}. This three-state BB84-like protocol is based on the complementarity approach~\cite{Koashi09} and uses the bases mismatch events information. Considering the imperfect phase/polarization modulation, it is demonstrated that the key rate is dramatically improved, although the phase error rate estimation technique is quite complicated.

The original BB84 protocol was described using  the qubits encoded into the polarization states of a single photon. In practice, the single-photon source is approximated by weak coherent pulses generated by a highly attenuated laser. There is a non-zero probability for such coherent state to contain multiple photons that can be exploited by Eve performing the photon-number-splitting (PNS) attack \cite{Brassard00,Lutkenhaus00}. The invention and further development of the decoy-state method \cite{Hwang03,Lo05,Wang05,Ma05} solves this problem by introducing extra states of different intensity (mean photon number per pulse) and significantly improves the GLLP bound on the maximum secure transmission distance. For a complete formal security proof of the decoy state method see, e.g., Ref.~\cite{Trushechkin21}. Due to inaccuracies of realistic intensity modulation, the intensities of the signal and decoy states can vary from pulse to pulse and, hence, cannot be considered as constant parameters of the protocol. It can open another potential loophole and allow Eve to improve her attack strategy. 

Another well known loophole of the leaky source is the Trojan-horse attack (THA) in which Eve injects bright light pulses into Alice’s device and then measures the back-reflected light in order to extract information about Alice’s state preparation like the basis choice and the intensity setting in the decoy-state QKD protocol. The information leakage from Alice's modulators during THA is studied in Refs.~\cite{Lucamarini15,Tamaki16,Wang18,Molotkov20,Molotkov21}. The THA on the loss-tolerant protocol is investigated in Refs.~\cite{Pereira19,Pereira20,Pereira23,Curras23} but only for the case of a single-photon source.

In this report we consider the most widely used two-decoy-state BB84 \cite{Ma05} with polarization encoding and focus on two types of source flaws -- the fluctuating coherent-state intensities and the polarization misalignment due to Alice's imperfect intensity and phase modulators. We assemble a simplified optical scheme for quantum state preparation that mimics the features of realistic commercial QKD devices and extract the intensity/phase distributions from laboratory test measurements. We study the non-Poissonian photon-number statistics following Refs.~\cite{Wang08,Wang09,Foletto22} in order to take into account the fluctuating signal and decoy pulse intensities. In contrast to the model-independent analysis in~\cite{Wang08,Wang09}, we do not rely on the knowledge of upper/lower bounds on intensities, but determine the intensity probability distributions directly from experiment. We extend this research and investigate also the imperfect encoding state preparation. Using the idea of fully-passive source~\cite{Wang23}, we compute the averaged density matrices and apply the \textit{imbalanced quantum coin} approach \cite{GLLP} to estimate the phase error rate. In particular, it is demonstrated for our setup that the encoding inaccuracies can lead to the key rate reduction of less than 50\% for channel losses up to 20\,dB (equivalent to transmission distances up to 100\,km), and the critical transmission distance can reach $\sim$120\,km. Since our experimental analysis is made for illustrative purposes only, we assume that the modulation precision and data analysis can be further improved, and the key rate bound can be increased. Provided with a complete set of explicit formulas, our framework can be applied to any practical QKD system with the characterized source that implements not the loss-tolerant but the usual decoy-state BB84 protocol.

This paper is organized as follows. In Section~\ref{sec:intensity}, we investigate the non-Poissonian photon-number statistics of weak coherent states and estimate the effect of pulse intensity fluctuations on the secret key rate. In Section~\ref{sec:phase}, we study the polarization distributions and bound the phase error rate for the QKD protocol with realistic phase modulator. The main conclusions are given in Section~\ref{sec:conclusions}.

\section{Intensity fluctuations}\label{sec:intensity}

\subsection{Experimental setup}

In this report, we investigate the efficient BB84 protocol \cite{Lo05a} with non-optimized basis choice probabilities of $p_X=0.9$ and $p_Y=0.1$ in which the $X$--basis is used for the secret key distillation while the $Y$--basis is used for the phase error rate estimation. In order to counteract the PNS attack, we apply the three-intensity decoy-state technique~\cite{Lo05,Ma05}: the signal state of intensity $\mu$ is used for the secret key generation, and two decoy states of intensities $\nu_1$ and $\nu_2$ are used for the single-photon yield and bit error rate estimation. For simplicity, we choose the reasonable values of $\mu\sim0.3$, $\nu_1\sim0.1$ and $\nu_2\sim10^{-3}$ which turn out to be close to the optimal ones for a wide range of transmission distances. The states are randomly generated with the probabilities of $p_\mu=0.5$ and $p_{\nu_{1,2}}=0.25$.

\begin{figure}[t!]\centering
	\includegraphics[width=0.9\textwidth]{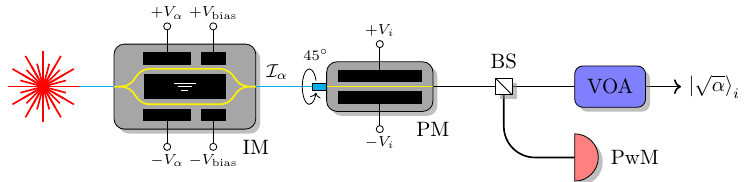}
	\caption{The optical scheme of weak coherent state preparation of given intensity and polarization: Light source -- DFB laser, IM -- intensity modulator  Optilab IMP-1550-10-PM, PM -- phase modulator IXblue MPZ-LN-10, BS -- beamsplitter with the split ratio of 1:99, PwM -- power meter Thorlabs PM100USB+S154C, VOA -- electronic variable optical attenuator. The cyan line denotes polarization-maintaining optical fiber. The connector to PM is rotated by 45$^\circ$ angle with respect to the PM crystal axes.}
	\label{fig:optic_scheme}
\end{figure}

Our experimental setup for preparing and monitoring Alice's quantum states of given intensity $\alpha\in\{\mu,\nu_1,\nu_2\}$ and polarization is presented in Fig.~\ref{fig:optic_scheme}. The pulses with an average optical power of approximately 2\,mW are emitted by a DFB laser operating at 1550\,nm wavelength. 
First, the relative classical pulse intensity $\I_\alpha$ is adjusted by the intensity modulator (IM) by applying an appropriate voltage $V_\alpha$ on the modulation electrodes such that $\I_{\nu_{1,2}}/\I_\mu=\nu_{1,2}/\mu$ with $\I_\mu$ chosen to be larger than half of the maximum output intensity. Then the polarization state is chosen by applying one of the four voltages $V_i$ on the phase modulator (PM). Finally, the classical pulses are attenuated to the required weak coherent-state level with the variable optical attenuator (VOA).

The IM output intensity $\I_\alpha$ is the result of the interference of two waves, propagating via two paths of the Mach-Zehnder interferometer with different optical path length. This difference is created by varying the optical index in the waveguide active layer of each path with electric field between the modulation electrodes. The IM is designed to have equal arms and thus balanced optical paths. However, there is always some imbalance, caused by the change of the refractive index of the optical mode of the waveguide due to possible variations of temperature (thermo-optic and pyroelectric effects), optical power in the waveguide (photorefractive effect) or mechanical stress (strain-optic effect). As a result, a time-dependent phase (and consequently $\I_\alpha$) drift is induced. In order to compensate this drift, the time-averaged power is measured by the power meter (PwM) and tuned by the proportional–integral–derivative (PID) controller to the initial value by applying the relevant voltage $V_\text{bias}$ to the bias electrodes. All these effects together with the limiting resolution of control voltage circuits and the laser noise cause some fluctuations of the output intensities.

\begin{figure}[t!]\centering
    \includegraphics[width=0.6\textwidth]{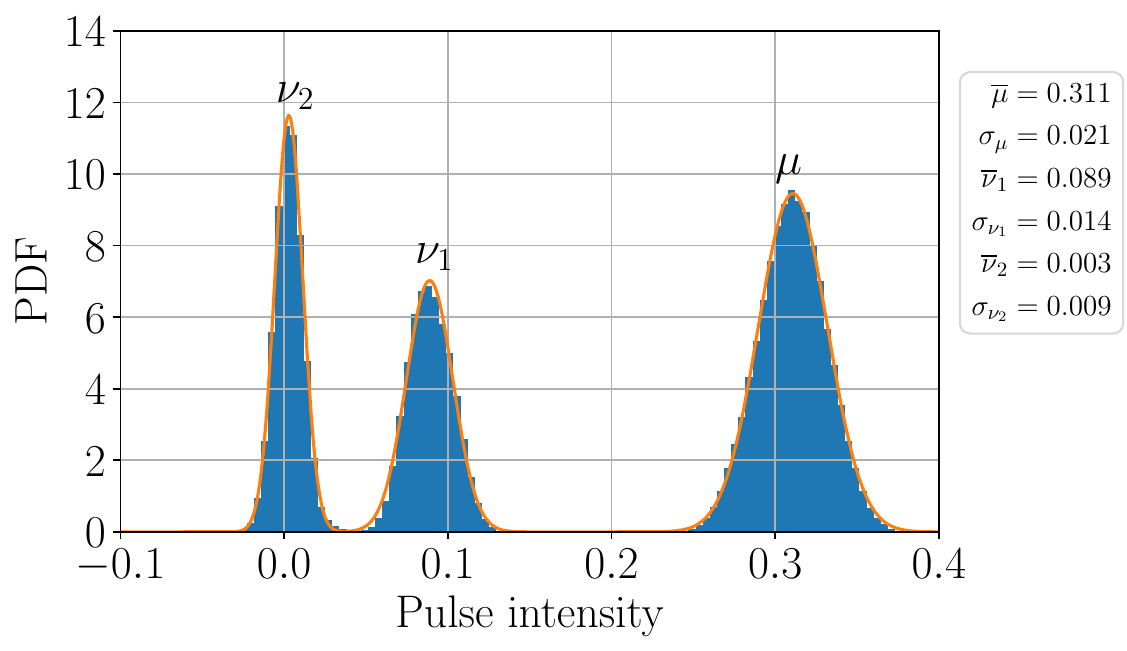}
    \caption{The measured probability density function of pulse intensity (mean photon number per pulse). The signal ($\mu$) and two decoy ($\nu_{1,2}$) states are generated randomly with the probabilities of $p_\mu=0.5$ and $p_{\nu_{1,2}}=0.25$ respectively. The non-physical negative values of the vacuum decoy peak are caused by photodetector noise.}
	\label{fig:hist_int}
\end{figure}

In order to quantify the intensity fluctuations, we connect a classical photodetector (Thorlabs RXM40AF) directly to the IM output and record the voltage oscillograms. Following Ref.~\cite{Huang23}, we remove the noise by applying filtering technique based on the singular-value-decomposition. In Fig.~\ref{fig:hist_int}, we present the obtained probability density function (PDF) of $\I_\alpha$ normalized by the total PM+BS+VOA attenuation. 
As evidenced by good agreement with the data, a regular Gaussian PDF given by Eq.~\eqref{eq:Gauss}, is the right choice of function to parametrize the intensity fluctuations :
\begin{equation}
    G(\alpha, \alphabar_i, \sigma_{\alpha_i}) = \frac{1}{ \sqrt{2 \pi}\sigma_{\alpha_i} } e^{-\frac{(\alpha - \alphabar_i)^2}{2{\sigma_{\alpha_i}^2}}} \,,
    \label{eq:Gauss}
\end{equation}
where $\alphabar_i$ are the mean values and $\sigma_{\alpha_i}$ are the standard deviations, with the best-fit values tabulated on the right of the plot in Fig.~\ref{fig:hist_int}.

\subsection{Non-Poissonian photon-number statistics}\label{sec:non-Poisson_stat}

The single-photon source is often approximated by a weak coherent-state source, realized in practice as strongly attenuated laser radiation. The emitted phase-randomized state is described as a mixture of Fock states,
\begin{equation}
    \rho_\alpha = \sum_{n=0}^\infty P_{n|\alpha} \ketbra{n}{n} = \sum_{n=0}^\infty e^{-\alpha} \frac{\alpha^n}{n!} \ketbra{n}{n} \,,
    \label{eq:rho_Poisson}
\end{equation}
where the photon number follows the Poisson distribution $P_{n|\alpha}$ with mean photon number $\alpha$. Note that usually in the literature $\alpha$ is assumed to be a constant parameter that does not vary during the QKD session. However, in realistic experimental setup the intensity parameter is a fluctuating variable as discussed in the previous subsection.

It is known that the assumption of Poissonian photon-number statistics is not really necessary for the decoy-state BB84 protocol -- one can derive the generalized security bounds for any arbitrary $P_{n|\alpha}$ distribution. The first detailed model-independent analysis of the source errors in the photon-number space is provided by Wang et al. in Refs.~\cite{Wang08, Wang09}. The authors derive the generalized conservative bounds on the single-photon component's yield and QBER in terms of arbitrary upper/lower bounds on $P_{n|\alpha}$. In this report, we closely follow the theoretical approach of \cite{Wang08, Wang09} but do not use the bounded $P_{n|\alpha}$ that rely on the knowledge of allowed intervals $[\alpha_{\min},\alpha_{\max}]$ for the coherent state source. Instead we assume $\rho_\alpha$ of every single pulse to be Poissonian but with random normally-distributed parameter $\alpha$. Thus, the $n$--photon state probability is modified as follows,
\begin{equation}
    P_{n|\alpha} = \frac{\int_0^\infty e^{-\alpha} \frac{\alpha^n}{n!} G(\alpha, \alphabar, \sigma_\alpha) d\alpha}{\int_0^\infty G(\alpha, \alphabar, \sigma_\alpha) d\alpha} \,,
    \label{eq:P_n}
\end{equation}
where we integrate only over the positive intensity values that have physical meaning.

The experimentally measured gain -- the probability that an emitted pulse of intensity $\alpha$ is detected by Bob -- can be expressed as
\begin{equation}
    Q_\alpha = \sum_{n=0}^\infty P_{n|\alpha} \Y_n \,,
\end{equation}
where the yield $\Y_n$ is the conditional probability of a detection event (click) on Bob’s side, given that Alice sends out an $n$--photon state. The key idea of the decoy-state method is that since Eve cannot distinguish the $\ket{n}$ states of signal pulse from those of decoy pulse, the yields $\{\Y_n\}_{n=0}^\infty$ must be equal for signal and decoy states~\cite{Hwang03}. However, this  assumption of indistinguishability may not hold in practice. In Ref.~\cite{Huang18}, the authors consider an imperfect source with passive side channels in the time and frequency domains which may allow Eve to partially distinguish signal and decoy states from the mismatches in the corresponding degrees of freedom. In this report, we neglect the impact of such side channels on the QKD system performance and focus only on the intensity fluctuations.

Taking into account the finite-statistics effects, the generalized lower bounds on the zero-photon and single-photon yields are determined by (for more details see Appendix~\ref{app:new_Y})
\begin{equation}
    \Y_0^l = \max\bigg\{ \frac{Q_{\nu_2}^l P_{1|\nu_1} - Q_{\nu_1}^u P_{1|\nu_2}}{P_{0|\nu_2} P_{1|\nu_1} - P_{0|\nu_1} P_{1|\nu_2}} ,\, 0 \bigg\} \,,
    \label{eq:Y0_l}
\end{equation}
\begin{equation}
    \Y_1^l = \frac{ Q_{\nu_1}^l P_{0|\nu_2} - Q_{\nu_2}^u P_{0|\nu_1} - \displaystyle\frac{P_{2|\nu_1} P_{0|\nu_2} - P_{2|\nu_2} P_{0|\nu_1}}{P_{2|\mu}} \big( Q_{\mu}^u - P_{0|\mu} \Y_0^l \big) }{P_{0|\nu_2} P_{1|\nu_1} - P_{0|\nu_1} P_{1|\nu_2}  - \displaystyle\frac{P_{2|\nu_1} P_{0|\nu_2} - P_{2|\nu_2} P_{0|\nu_1}}{P_{2|\mu}} P_{1|\mu}} \,.
    \label{eq:Y1_l}
\end{equation}
where $Q_\alpha^{u,l}$ are the upper/lower bounds on the $Q_\alpha$--estimators given by Eq.~\eqref{eq:Q_ul}. For brevity, here we omit the basis index of $Q_\alpha^{u,l}$ and $\Y_{0,1}^l$.

\begin{figure}[t!]\centering
	\includegraphics[width=0.49\textwidth]{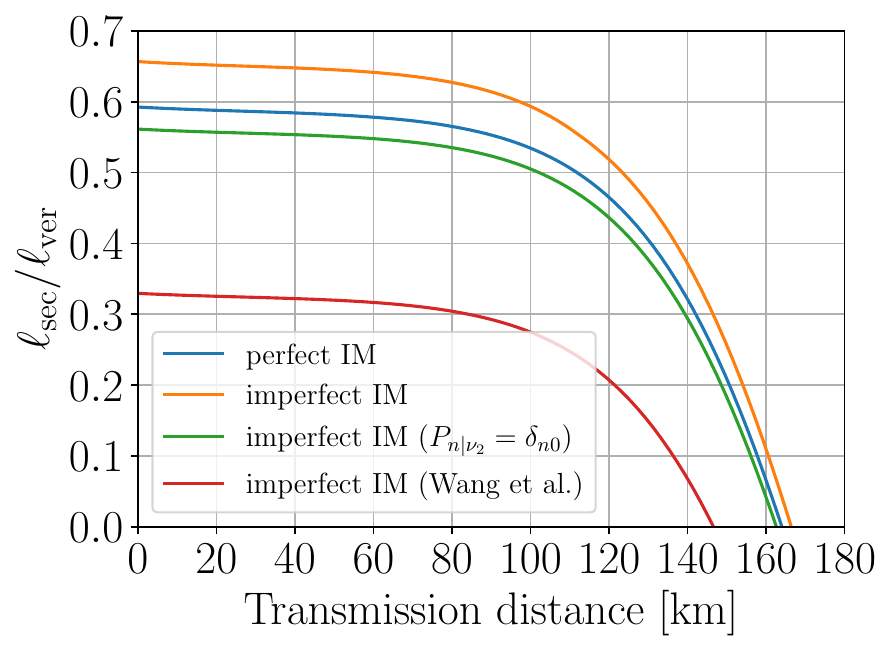}
    \caption{The simulated secret key length $\ell_\sec$, normalized to the verified key length $\ell_\ver$, for the scenarios of perfect ($\mu,\nu_1,\nu_2$ are constant) and imperfect ($\mu,\nu_1,\nu_2$ fluctuate following the normal distribution) pulse intensity modulation. The red curve represents the result of the method introduced by Wang et al. in \protect\cite{Wang08,Wang09} with the intensities bounded with their $\pm1\sigma$ uncertainties.}
	\label{fig:r_sec_IM}
\end{figure}

To investigate the effect of non-Poissonian statistics on the privacy amplification, in Fig.~\ref{fig:r_sec_IM} we plot the simulated ratio of the secret over verified key lengths, $\ell_\sec/\ell_\ver$, as a function of transmission distances for the scenarios of perfect (Poissonian statistics with $P_{n|\alpha}=e^{-\alphabar}\,\alphabar^n/n!$) and imperfect (non-Poissonian statistics with $P_{n|\alpha}$ defined by Eq.~\eqref{eq:P_n}) intensity modulation. The verified key is the error-corrected sifted key that passed the subsequent hash-tag verification. The secret key length formula is given in Appendix~\ref{app:l_sec}. Here we assume the photon source to be basis-independent ($\Delta^\prime=0$ in Eq.~\eqref{eq:Theta_coin}). One can see from the plot that, surprisingly, the imperfect preparation provides a better key rate than the perfect one for the entire distance range. The reason is the following: due to rather wide distribution of $\nu_2$ relative to its mean value, the probability of the single-photon component of $\rho_{\nu_2}$ turns out to be enhanced several times with respect to the ideal Poissonian source with the mean photon number $\overline\nu_2$; as a consequence, the term $P_{0|\nu_1}P_{1|\nu_2}$ in the denominator of Eq.~\eqref{eq:Y1_l} becomes non-negligible, which, in turn, increases $\Y_1^l$ and, hence, $\ell_\sec$. One could naively think that increasing $\sigma_{\nu_2}$ would improve $\ell_\sec$ even more, however, this would lead to the violation of the required condition~\eqref{eq:check1} starting from some value of $n$, making the $\Y_0$--estimation \eqref{eq:Y0_derivation} incorrect. Increasing $\sigma_{\nu_2}$, e.g., by 5 times would violate \eqref{eq:check1} for $n\geq12$. In order to avoid the overestimation of the single-photon and multi-photon contributions to $\rho_{\nu_2}$ due to the noise, we propose to set $P_{n|\nu_2}=\delta_{n0}$ (i.e. $\nu_2\equiv0$ and $\rho_{\nu_2}=\ketbra{0}{0}$) which provides a more conservative evaluation of $\ell_\sec$ (see the green curve in Fig.~\ref{fig:r_sec_IM}). In this case the secret key turns out to be more than 95\% of the key generated with an ideal IM, for any distance up to 100\,km.

To conclude this section, we make a comparison with the model-independent approach of Wang et al.~\cite{Wang08,Wang09}. Their expression for the single-photon gain, given by Eq.~(43) in Ref.~\cite{Wang09}, can be re-written in our notations as
\begin{equation}
    Q_1^l = P_{1|\mu}^l \Y_1^l =
    P_{1|\mu}^l \frac{Q_{\nu_1}^l P_{2|\mu}^l - Q_\mu^u P_{2|\nu_1}^u - \big(P_{2|\mu}^l P_{0|\nu_1}^u - P_{2|\nu_1}^u P_{0|\mu}^l\big) \displaystyle{\frac{Q_{\nu_2}^u}{P_{0|\nu_2}^l}}}{P_{2|\mu}^l P_{1|\nu_1}^u - P_{2|\nu_1}^u P_{1|\mu}^l} \,,
    \label{eq:Q1_l_Wang}
\end{equation}
with the bounded probabilities for coherent states defined as
\begin{equation}
    P_{n|\alpha}^{u(l)} =
    \begin{cases}
        e^{-\alpha^{l(u)}} \,, &\quad n=0 \\
        \frac{(\alpha^{u(l)})^n}{n!} e^{-\alpha^{u(l)}} \,, &\quad n=1,2
    \end{cases}
\end{equation}
Using the experimental intensity distributions shown in Fig.~\ref{fig:hist_int}, we determine $\alpha^{u,l}$ as
\begin{equation}
    \alpha^{u(l)} = \alphabar \pm z_{1-\varepsilon} \sigma_\alpha \,, \quad \Pr(\alpha \leq \alpha^l) = \Pr(\alpha \geq \alpha^u) = \varepsilon \,,
\end{equation}
where $z_{1-\varepsilon}$ is the normal distribution quantile,
\begin{equation}
	z_{1-\varepsilon} = \Phi^{-1} (1 - \varepsilon) = \sqrt2 \,{\rm erf}^{-1}(1 - 2\varepsilon) \,.
\end{equation}
For illustration, the red curve in Fig.~\ref{fig:r_sec_IM} represents the normalized secret key length~\eqref{eq:l_sec} computed with $Q_1^l$ defined by Eq.~\eqref{eq:Q1_l_Wang} and $z_{1-\varepsilon}=1$ ($\varepsilon\simeq0.16$). We find that for $z_{1-\varepsilon}\gtrsim2.3$ ($\varepsilon\lesssim0.011$) the secret key transmission is impossible at any distance. The wrong estimation of at least one bound on $P_{n|\alpha}$ in \eqref{eq:Q1_l_Wang} would make the evaluation of $Q_1^l$ and, hence, $\ell_\sec$ incorrect. Therefore, such failure probability of $1-(1-\varepsilon)^5\approx5\varepsilon$ has to be taken into account in the total tolerable failure probability $\varepsilon_\text{decoy}$ which is set to be $10^{-12}$ (see Appendix~\ref{app:l_sec}). Apparently, such high confidence level cannot be achieved with the current values of $\sigma_\alpha$. Thus, one can clearly see superiority of the proposed method for the particular setup and measurement precision since it provides both longer secret keys and the maximum attainable transmission distances.

\section{Phase fluctuations}\label{sec:phase}

\subsection{Polarization state preparation and basis-dependence}

Any arbitrary polarization state can be described by azimuthal and polar angles $\varphi\in[0,2\pi)$ and $\theta\in[0,\pi]$ on the Bloch sphere,
\begin{equation}
	\ket{\psi(\varphi,\theta)} = \cos\bigg(\frac{\theta}{2}\bigg) \ket{H} + e^{i\varphi} \sin\bigg(\frac{\theta}{2}\bigg) \ket{V} \,,
\end{equation}
where $\ket{H}$ and $\ket{V}$ denote the horizontal and vertical polarization vectors, aligned with the PM crystal axes. In our setup, the linearly polarized light is injected to PM at a 45$^\circ$ angle to the crystal axes which corresponds to $\theta=\pi/2$ \cite{Duplinskiy17}. In general, the electric field amplitudes along the ordinary and extraordinary axes initially have some phase difference $\phi_0$ without any voltage supply due to the crystal birefringence. The additional random relative phase $\phi_i\in\{0,\pi,\pi/2,3\pi/2\}$, determining the basis and bit value, is created by applying an appropriate voltage $V_i$ along one of the crystal axes. Thus, the azimuthal angle of the $i^\text{th}$ state is $\varphi_i=\phi_0+\phi_i$.

\begin{figure}[t!]\centering
	\includegraphics[width=0.45\textwidth]{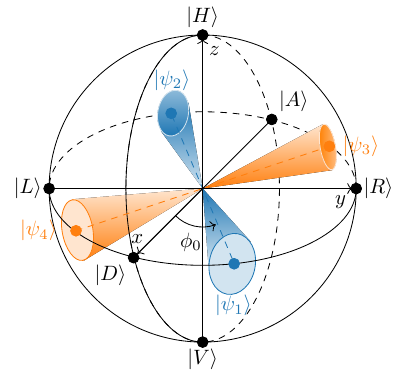}
	\caption{Alice's output polarization states on the Bloch sphere. The blue and orange dots mark the perfectly prepared states \eqref{eq:psi_ideal}, while the realistic states \eqref{eq:psi_real} are schematically depicted as solid angle surface areas.}
	\label{fig:Poincare_sphere}
\end{figure}

In the protocol with perfect state preparation, the qubits are prepared and measured in the elliptical polarization bases $\Xp:\{\ket{\psi_1},\ket{\psi_2}\}$ and $\Yp:\{\ket{\psi_3},\ket{\psi_4}\}$, obtained by rotating the standard bases $X:\{\ket{D},\ket{A}\}$ and $Y:\{\ket{R},\ket{L}\}$ around the $z$--axis by $\phi_0$ (see Fig.~\ref{fig:Poincare_sphere}), with the basis vectors in the form of
\begin{equation}
	\ket{\psi_{1,2}^\text{perfect}} = \frac{1}{\sqrt2} \big( \ket{H} \pm e^{i\phi_0} \ket{V} \big) \,, \quad
    \ket{\psi_{3,4}^\text{perfect}} = \frac{1}{\sqrt2} \big( \ket{H} \pm i e^{i\phi_0} \ket{V} \big) \,.
	\label{eq:psi_ideal}
\end{equation}
Then Bob performs his measurement in the $\Xp$ and $\Yp$ bases by randomly applying one of the positive operator-valued measures $\{\ketbra{\psi_i^\text{perfect}}{\psi_i^\text{perfect}}\}$.

However, in practice, the light enters PM not ideally at 45$^\circ$ due to mechanical inaccuracy of connection between the optical components. This error induces a deviation from $\theta=\pi/2 $ on the Bloch sphere. The voltage control and, hence, $\{\phi_i\}$ have some uncertainty as well. As a consequence of all these imperfections, the ideal states \eqref{eq:psi_ideal} float above/below the $xy$--plane and rotate around the $z$--axis, fluctuating around their average positions. So the physical states sent to Bob can be written as
\begin{equation}
	\ket{\psi_i} = \ket{\psi(\varphi_i, \theta)} \,,
	\label{eq:psi_real}
\end{equation}
where $\varphi_i$ is a random variable with some probability distribution while $\theta$ is an \textit{a priori} unknown that is determined from experiments with some uncertainty. Note that, in general, these states are no longer mutually orthogonal in the corresponding basis and do not lie in the $xy$--plane.

Using the density matrix formalism, the $\Xp$--basis and $\Yp$--basis states are described by $\rho_\Xp=\frac{1}{2}(\ketbra{\psi_1}{\psi_1}+\ketbra{\psi_2}{\psi_2})$ and $\rho_\Yp=\frac{1}{2}(\ketbra{\psi_3}{\psi_3}+\ketbra{\psi_4}{\psi_4})$ that can be written as
\begin{equation}
    \begin{split}
        \rho_\Xp &= \frac{1}{2}
        \begin{pmatrix}
            1+\cos\theta & e^{-i\frac{\varphi_1+\varphi_2}{2}} \cos\big(\frac{\varphi_1-\varphi_2}{2}\big) \sin\theta \\
            e^{i\frac{\varphi_1+\varphi_2}{2}} \cos\big(\frac{\varphi_1-\varphi_2}{2}\big) \sin\theta & 1-\cos\theta
        \end{pmatrix} \,, \\
        \rho_\Yp &= \frac{1}{2}
        \begin{pmatrix}
            1+\cos\theta & e^{-i\frac{\varphi_3+\varphi_4}{2}} \cos\big(\frac{\varphi_3-\varphi_4}{2}\big) \sin\theta \\
            e^{i\frac{\varphi_3+\varphi_4}{2}} \cos\big(\frac{\varphi_3-\varphi_4}{2}\big) \sin\theta & 1-\cos\theta
        \end{pmatrix} \,.
    \end{split}
    \label{eq:rho_XY}
\end{equation}
For the perfectly prepared states \eqref{eq:psi_ideal} $\rho_\Xp=\rho_\Yp$, i.e. the photon source is basis-independent. If $\rho_\Xp\neq\rho_\Yp$ one cannot simply estimate the unknown single-photon phase error rate in the $\Xp$--basis $E_1^{\ph,\Xp}$ by the measured bit error rate in the $\Yp$--basis $E_1^\Yp$ (or vice versa). The more state dependence on the basis, the easier for Eve to distinguish the bases and hence the lower secret key rate.

In Ref.~\cite{Wang23}, dedicated to the fully-passive QKD, the authors consider an equivalent virtual entanglement-based protocol
with a source that emits perfectly encoded pure decoy states in the $X,Y$--bases and signal states with mixed polarizations in the $Z$--basis. It is shown that Alice's imperfect preparation in the $Z$--basis is equivalent to the imperfect measurement (i.e. the trusted noise in Alice’s post-processing) and does not affect the amount of privacy amplification. However, the authors make an assumption that the polarization fluctuations in the $Z$--basis are symmetric on the Bloch sphere. It implies that averaged over angles $\rho_Z$ is the fully-mixed state, i.e. $\rhobar_H+\rhobar_V=\mathbb{I}$ where
\begin{equation}
    \begin{split}
        \rhobar_H &= \int_0^{2\pi} \int_0^{\delta\theta} p(\varphi,\theta) \ketbra{\psi(\varphi,\theta)}{\psi(\varphi,\theta)} d\varphi d\theta \,, \\
        \rhobar_V &= \int_0^{2\pi} \int_{\pi-\delta\theta}^\pi p(\varphi,\theta) \ketbra{\psi(\varphi,\theta)}{\psi(\varphi,\theta)} d\varphi d\theta \,,
    \end{split}
\end{equation}
are the mixed states, post-selected within two cones around the $z$--axis with half-angle $\delta\theta$, with symmetric probability distribution $p(\varphi,\theta)=p(\varphi+\pi,\theta)$.

This approach is not applicable in our case since we do not assume any symmetry of angular probability distributions $\{p_i(\varphi,\theta)\}$. Nevertheless, we can use one of the ideas of Refs.~\cite{Wang23},\cite{Zapatero23} -- the replacement of the source of randomly fluctuating pure states $\{\ket{\psi_i}\}$ by an equivalent source emitting the mixed states $\{\rhobar_i\}$,
\begin{equation}
    \rhobar_i = \int_0^{2\pi} \int_0^\pi p_i(\varphi, \theta) \ketbra{\psi(\varphi, \theta)}{\psi(\varphi, \theta)} d\varphi d\theta \,,
    \label{eq:mixed}
\end{equation}
where $\{p_i(\varphi,\theta)\}$ are extracted directly from experiment. In this case, the density matrices \eqref{eq:rho_XY} are substituted for
\begin{equation}
    \rhobar_\Xp = \rhobar_1 + \rhobar_2 \,, \quad
    \rhobar_\Yp = \rhobar_3 + \rhobar_4 \,.
    \label{eq:rho_XY_averaged}
\end{equation}

To quantify the discrepancy between $E_1^{\ph,\Xp}$ and $E_1^\Yp$, we use the concept of \textit{quantum coin}, introduced in Ref.~\cite{GLLP} in the equivalent virtual entanglement-based protocol in which both Alice and Bob measure in one basis and Alice announces the other basis. Applying the complementarity argument \cite{Koashi09} and the Bloch sphere bound \cite{Tamaki03} to the quantum coin yields the following inequality \cite{Lo07},
\begin{equation}
    \sqrt{F(\rhobar_\Xp, \rhobar_\Yp)} \leq 1 - \Y_1 + \Y_1 \bigg(\sqrt{E_1^{\ph,\Xp} E_1^\Yp} + \sqrt{(1 - E_1^{\ph,\Xp})(1 - E_1^\Yp)}\bigg) \,,
    \label{eq:F_inequality}
\end{equation}
with $\Y_1=(\Y_1^\Xp+\Y_1^\Yp)/2$ and \textit{fidelity} $F$ between two states, defined as
\begin{equation}
	F(\rhobar_\Xp, \rhobar_\Yp) \equiv \bigg[\text{Tr}\bigg(\sqrt{\sqrt{\rhobar_\Xp} \, \rhobar_\Yp \, \sqrt{\rhobar_\Xp}}\bigg)\bigg]^2 \,.
    \label{eq:F_alt}
\end{equation}
For $2\times2$ matrices it is more convenient to use the simplified form \cite{Jozsa94},
\begin{equation}
    F(\rhobar_\Xp, \rhobar_\Yp) = \text{Tr}\big(\rhobar_\Xp \, \rhobar_\Yp\big)+2\sqrt{\det(\rhobar_\Xp)\det(\rhobar_\Yp)} \,.
\end{equation}
Solving \eqref{eq:F_inequality}, one obtains the following upper bound on the single-photon phase error rate,
 \begin{equation}
    E_1^{\ph,\Xp} \leq E_1^\Yp + 4 \Delta^\prime (1 - \Delta^\prime) (1 - 2E_1^\Yp) + 4(1 - 2\Delta^\prime) \sqrt{\Delta^\prime (1 - \Delta^\prime) E_1^\Yp (1 - E_1^\Yp)} \,,
    \label{eq:E1_coin_alt}
\end{equation}
\begin{equation}
    \Delta^\prime = \frac{1 - \sqrt{F(\rhobar_\Xp, \rhobar_\Yp)}}{2\Y_1} = \frac{\Delta}{\Y_1} \,,
\end{equation}
where the quantity of $\Delta=(1-\sqrt{F})/2$ is usually called the \textit{quantum coin imbalance}. The lower and upper bounds on $\Y_1^{\Xp,\Yp}$ and $E_1^\Yp$ are determined via decoy-state method and are given by Eqs.~\eqref{eq:Y1_l} and \eqref{eq:E1_u}, respectively. The finite-key-size effects are taken into account in the modified $E_1^{\ph,\Xp}$ formula \eqref{eq:E1ph_finite-key}.

\subsection{Experimental data analysis}

\begin{figure}[t!]\centering
    \includegraphics[width=0.49\textwidth]{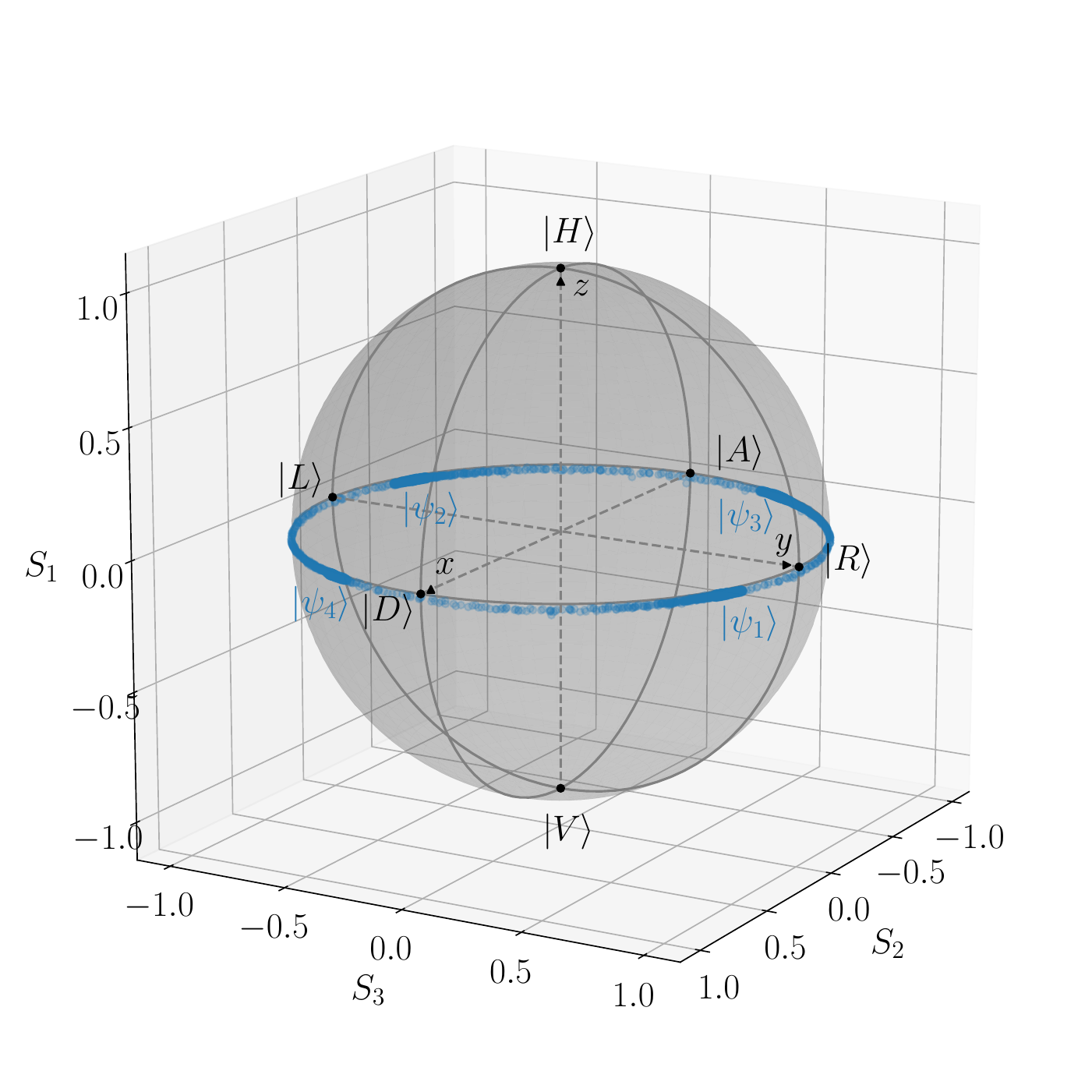}
    \includegraphics[width=0.49\textwidth]{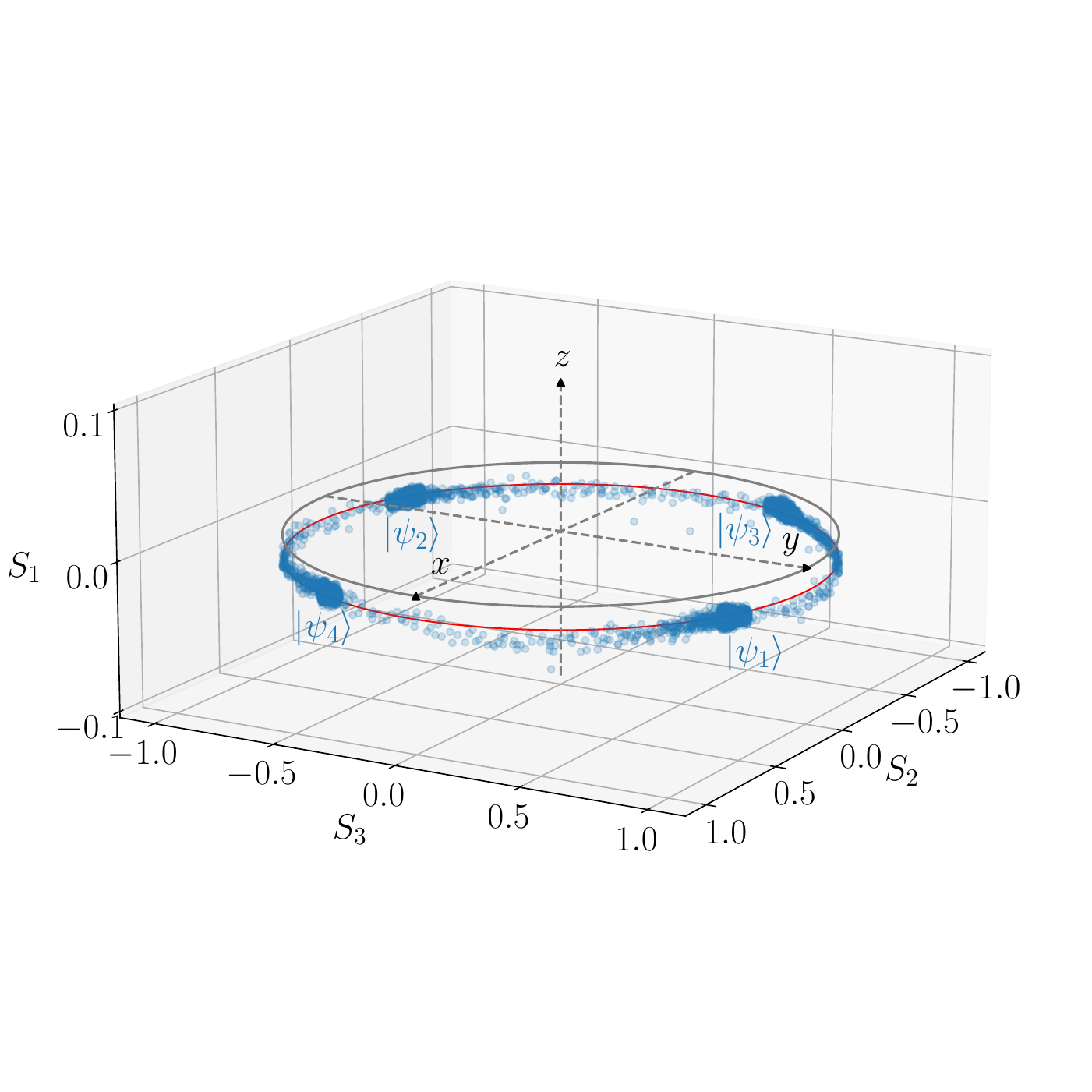}
	\caption{The measured polarization states \eqref{eq:psi_real} on the Poincar\'{e} sphere, made in the Stokes parameter coordinates $(S_1,S_2,S_3)$. The plot on the right shows the zoomed equatorial area of the sphere. The red circle is the best fit circle given 3D points.}
	\label{fig:Stokes}
\end{figure}

Using the optical scheme shown in Fig.~\ref{fig:optic_scheme}, we perform a series of polarization state measurements with the fast polarimeter (model PSY-201 by General Photonics). A periodic sequence of repeated constant voltage pulses corresponding to the phase retardation angles of $\{0,\pi/2,\pi,3\pi/2\}$ is applied to the phase modulator. The time duration of each pulse is $\sim52\,\mu$s. A representative result of such polarization measurements in terms of three-component Stokes vector $(S_1,S_2,S_3)$ is shown in Fig.~\ref{fig:Stokes}. One can see that the data points form a ring which is slightly shifted below the primary $xy$--plane of the Poincar\'{e} sphere. The majority of individual data points are grouped into four primary spots that are turned on that ring at an angle of $\phi_0\sim60^\circ$ around the $z$--axis of the sphere (see Fig.~\ref{fig:Stokes}, right panel). Sparse points between the primary four dense regions are artifacts of not high enough acquisition rate of the polarimeter. 

\begin{figure}[t!]\centering
    \includegraphics[width=0.43\textwidth]{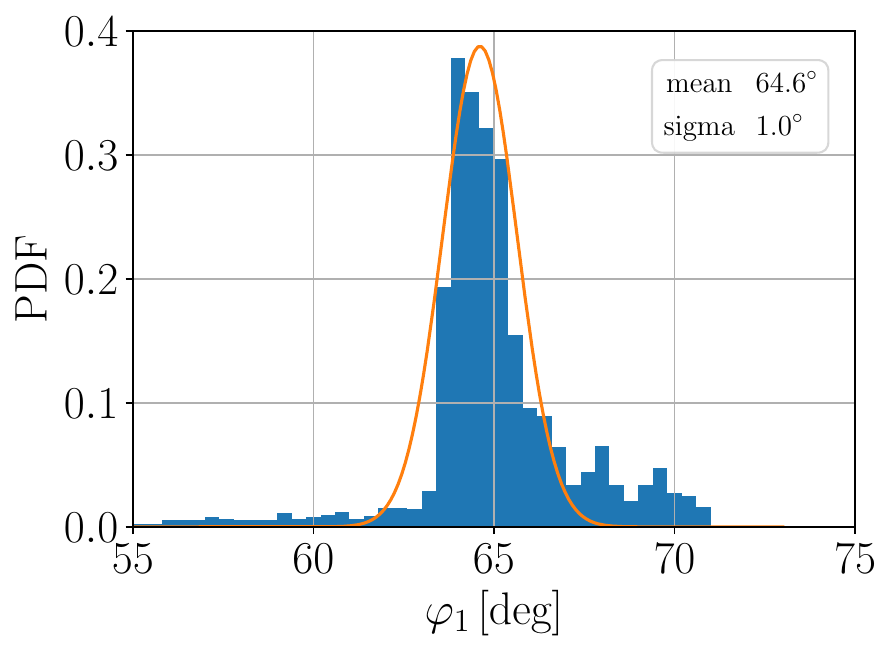}
    \includegraphics[width=0.43\textwidth]{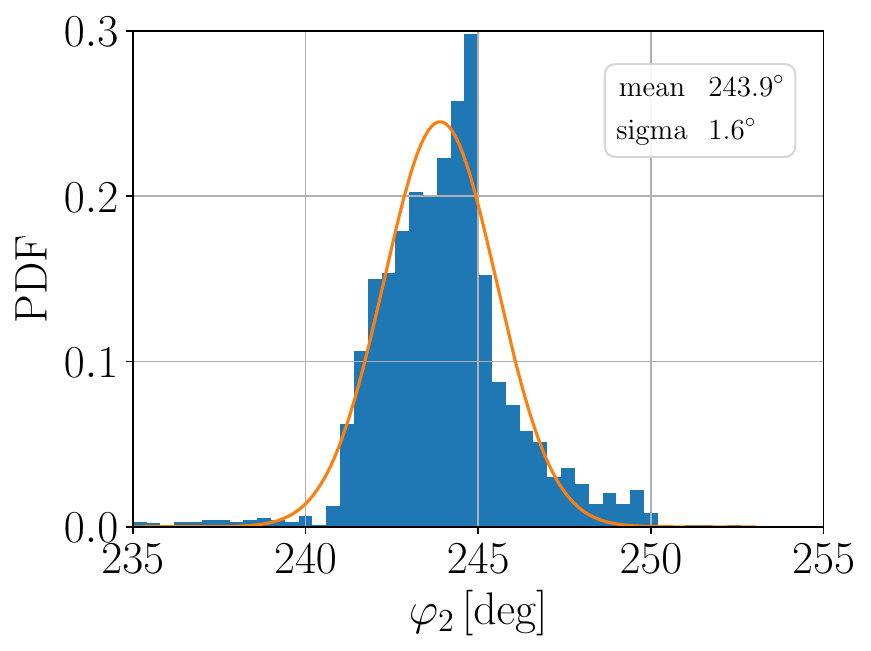}
    \includegraphics[width=0.43\textwidth]{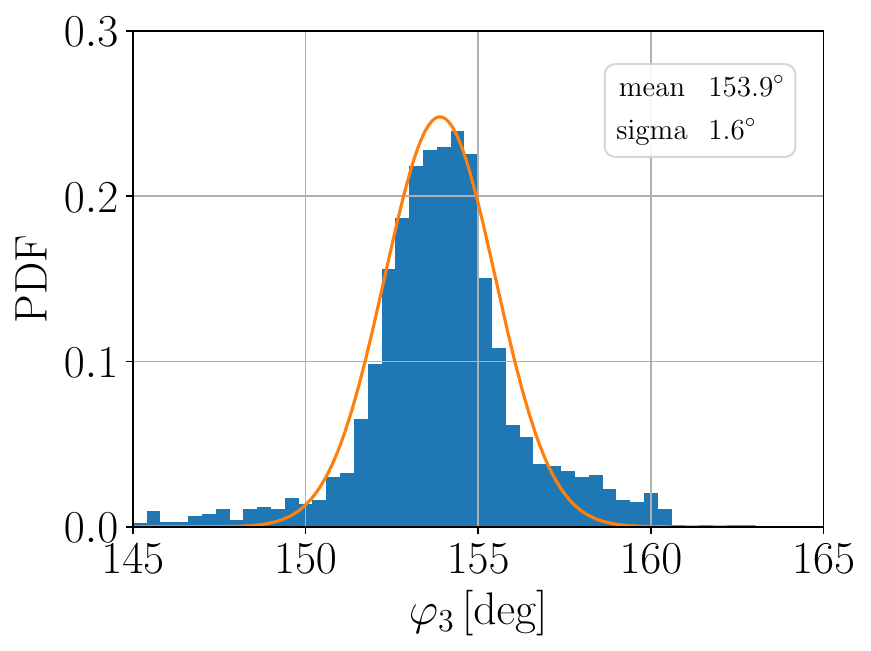}
    \includegraphics[width=0.43\textwidth]{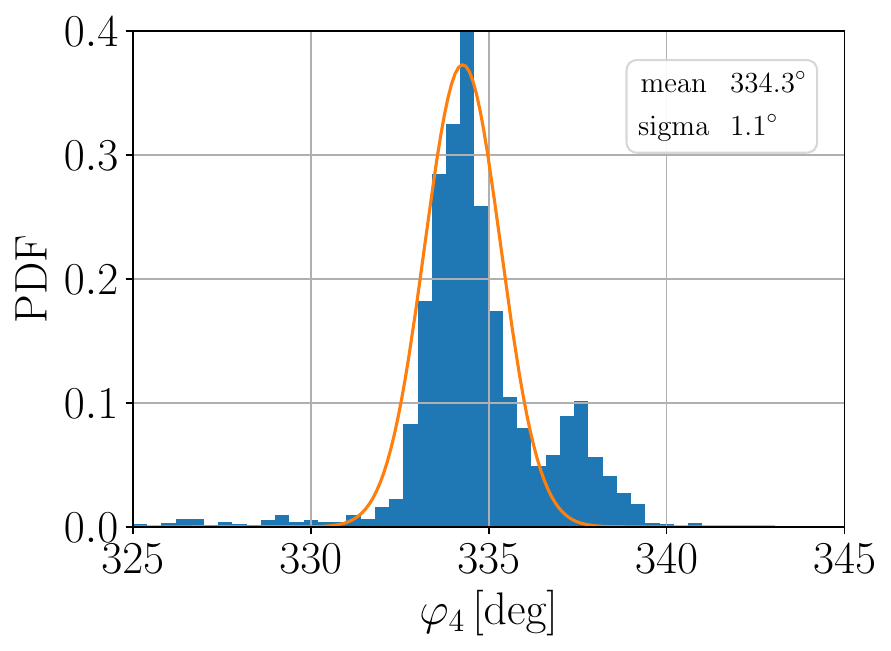}
    \includegraphics[width=0.43\textwidth]{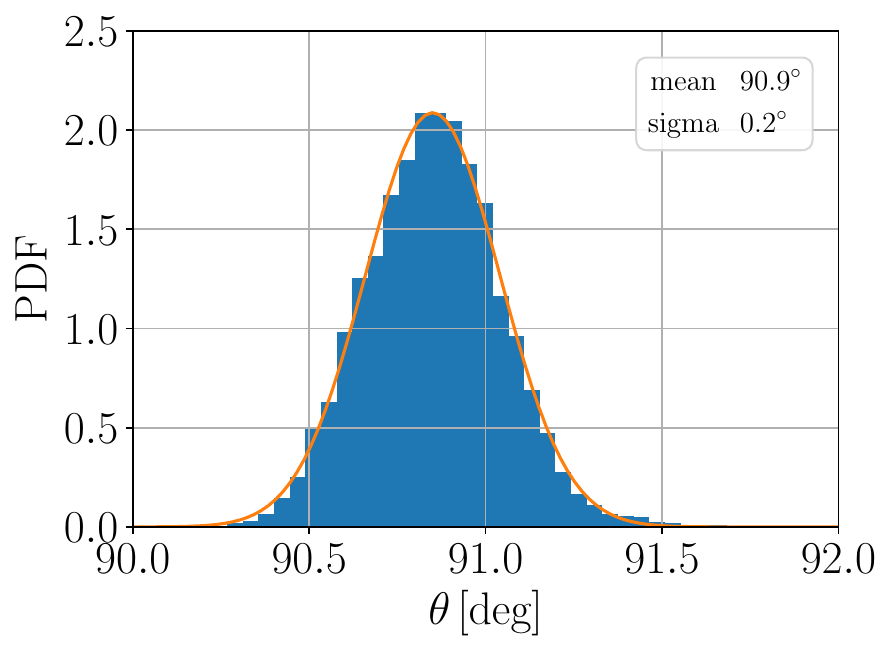}
	\caption{The experimental angular distributions, obtained from the spherical coordinates of the data points on the Poincar\'{e} sphere in Fig.~\ref{fig:Poincare_sphere}.}
	\label{fig:angle_histos}
\end{figure}

\begin{figure}[t!]\centering
	\includegraphics[width=0.49\textwidth]{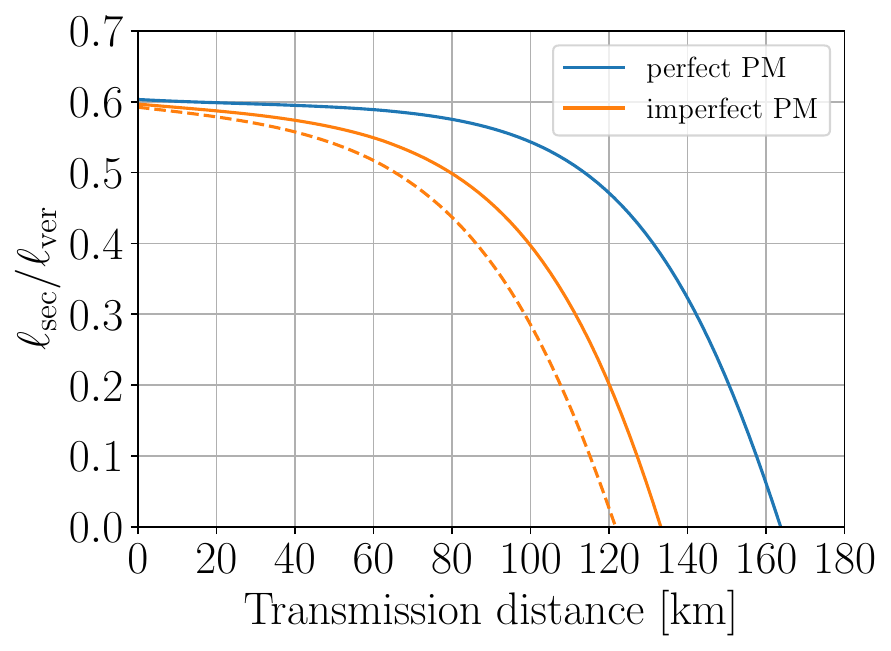}
    \includegraphics[width=0.49\textwidth]{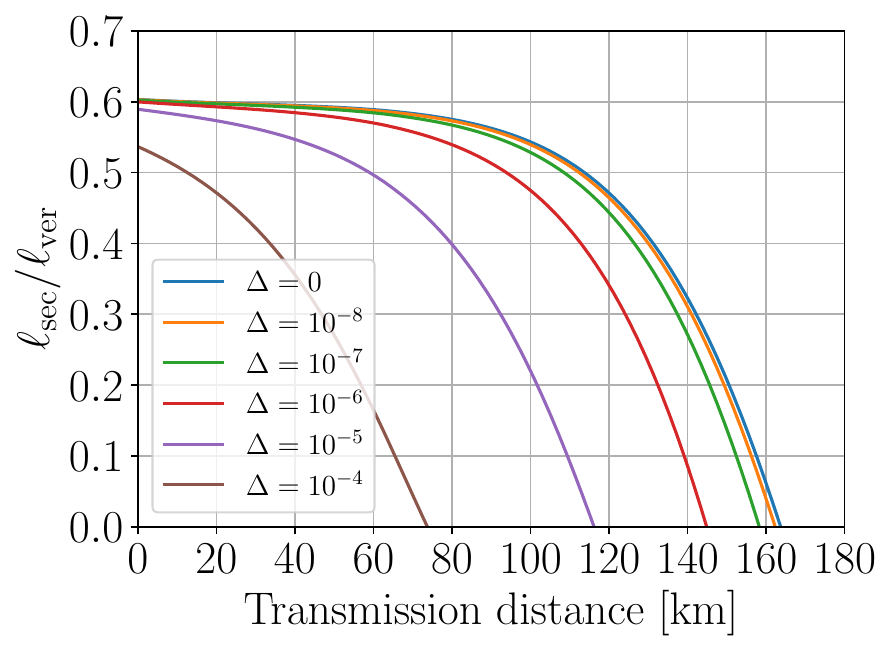}
    \caption{Left: The simulated secret key length $\ell_\sec$, normalized to the verified key length $\ell_\ver$, for scenarios of perfect and imperfect phase modulation. The orange solid(dashed) line corresponds to the fitted Gaussian (experimental binned) probability density function of angular distributions. Right: The simulated $\ell_\sec/\ell_\ver$ ratio for various values of quantum coin imbalance $\Delta$.}
	\label{fig:r_sec_PM}
\end{figure}

The typical experimental distributions of azimuthal ($\varphi$) and polar ($\theta$) angles,
\begin{equation}
    \varphi = \text{sgn}(S_3) \arccos\frac{S_2}{\sqrt{S_2^2 + S_3^2}} \,, \quad
    \theta = \arccos\frac{S_1}{\sqrt{S_1^2 + S_2^2 + S_3^2}} \,,
\end{equation}
are presented in Fig.~\ref{fig:angle_histos}.
One can notice that the $\varphi_i$--distributions are clearly asymmetric, however, the study of this asymmetry origin is beyond the scope of this report. Therefore, first, we make a conservative estimation of $\Delta$ using the model-independent approach: we extract the experimental normalized binned PDFs $p_{\varphi_i}^\exp$ and $p_\theta^\exp$ from Fig.~\ref{fig:angle_histos} and compute $\rhobar_i$ as follows,
\begin{equation}
    \rhobar_i = \sum_{n,m} p_{\varphi_i}^\exp(\varphi_i^{\bin\,n}) p_\theta^\exp(\theta^{\bin\,m}) \ketbra{\psi(\varphi_i^{\bin\,n},\theta^{\bin\,m})}{\psi(\varphi_i^{\bin\,n},\theta^{\bin\,m})} w_{\varphi_i} w_\theta \,,
\end{equation}
where $x^{\bin\,n}$ is the center of $n^\text{th}$ histogram bin of width $w_x$ of measured $x$--distribution ($x\in\{\varphi_i,\theta\}$). Then for the density matrices computed in this way we obtain $\Delta=7\times10^{-6}$. In Fig.~\ref{fig:r_sec_PM} we plot the privacy amplification factor $\ell_\sec/\ell_\ver$ where $\ell_\sec$ is determined by Eq.~\eqref{eq:l_sec} with corrected single-photon phase error rate~\eqref{eq:E1_coin_alt}. Here we assume no intensity fluctuations (i.e. $P_{n|\alpha}=e^{-\alphabar}\,\alphabar^n/n!$ and $\Y_{0,1}^l$ are determined by Eqs.~\eqref{eq:Y0_l_Poisson},\eqref{eq:Y1_l_Poisson}). One can see from the plot that that the critical distance at which the secret key generation becomes impossible is reduced by approximately 40\,km with respect to the perfect phase modulation case. One can also remark that the secret key rate reduction is up to 9 and 47\% for the short-distance (up to 50\,km) and long-distance (up to 100\,km) ranges, respectively.

We assume that increasing the data statistics and improving the measurement procedure with subsequent data post-processing will make the distributions smoother and symmetric. Applying a simple Gaussian model, we obtain for future prospects the following analytical approximation (verified by numerical integration),
\begin{equation}
    \begin{split}
        \rhobar_i &\simeq \int_0^{2\pi} \int_0^\pi G\big(\varphi, \phibar_i, \sigma_{\varphi_i}\big) G\big(\theta, \thetabar, \sigma_\theta\big) \ketbra{\psi(\varphi, \theta)}{\psi(\varphi, \theta)} d\varphi d\theta \\
        &\simeq \frac{1}{2}
        \begin{pmatrix}
        1 + e^{-\frac{\sigma_\theta^2}{2}} \cos\thetabar
        & e^{-i\phibar_i - \frac{1}{2} (\sigma_{\varphi_i}^2 + \sigma_\theta^2)} \sin\thetabar \\ 
        e^{i\phibar_i - \frac{1}{2} (\sigma_{\varphi_i}^2 + \sigma_\theta^2)} \sin\thetabar
        & 1 - e^{-\frac{\sigma_\theta^2}{2}} \cos\thetabar
        \end{pmatrix} \,,
    \end{split}
	\label{eq:rho_i_averaged}
\end{equation}
where for simplicity we omit the properly defined PDF normalization factors since
\begin{equation}
    \int_0^{x_{\max}} G(x,\overline{x},\sigma_x)dx = \frac{1}{2}\bigg[\erf\bigg(\frac{\overline{x}}{\sqrt2\sigma}\bigg) + \erf\bigg(\frac{x_{\max} - \overline{x}}{\sqrt2\sigma}\bigg)\bigg] \simeq 1 \,,
\end{equation}
if $\sigma_x\ll\overline{x}$ and $\sigma_x\ll x_{\max}-\overline{x}$. For the same reason, we replace the integration limits in~\eqref{eq:rho_i_averaged} by $\pm\infty$ and obtain a very simple analytical matrix expression above. Using the fitted parameter values of $\{\phibar_i,\sigma_{\varphi_i},\thetabar,\sigma_\theta\}$ from Fig.~\ref{fig:angle_histos}, we compute the fidelity between $\rhobar_\Xp$ and $\rhobar_\Yp$~\eqref{eq:F_alt} and find $\Delta=3\times 10^{-6}$. One can see from the plot in Fig.~\ref{fig:r_sec_PM} that the critical distance is reduced by approximately 30\,km compared to the ideal modulation case. Also one can note that for the short(long)-distance range the secret key length is reduced by less than 5(27)\%. These results turn out to be less conservative than those obtained with the model-independent approach, providing better secret key generation rate.

Finally, we also compare with a very conservative estimation of $\Delta$ by finding the minimum of fidelity (i.e. maximum of $\Delta$),
\begin{equation}
    \begin{split}
        F_{\min} &= \min_{\substack{\varphi_i \in [\varphi_i^{\min}, \varphi_i^{\max}] \\ \theta \in [\theta^{\min}, \theta^{\max}] }} F(\rho_\Xp, \rho_\Yp) =
        \min_{\substack{\varphi_i \in [\varphi_i^{\min}, \varphi_i^{\max}] \\ \theta \in [\theta^{\min}, \theta^{\max}] }}
        \frac{1}{2} \bigg\{ 1 + \cos^2\theta \\
        &+ \cos\bigg(\frac{\varphi_1 + \varphi_2 - \varphi_3 - \varphi_4}{2}\bigg) \cos\bigg(\frac{\varphi_1 - \varphi_2}{2}\bigg) \cos\bigg(\frac{\varphi_3 - \varphi_4}{2}\bigg) \sin^2\theta \\
        &+ \bigg|\sin\bigg(\frac{\varphi_1 - \varphi_2}{2}\bigg) \sin\bigg(\frac{\varphi_3 - \varphi_4}{2}\bigg)\bigg| \sin^2\theta \bigg\} \,,
    \end{split}
\end{equation}
with $\rho_\Xp$ and $\rho_\Yp$, defined by Eq.~\eqref{eq:rho_XY}. The minimization ranges are determined from the experimental distributions in Fig.~\ref{fig:angle_histos} as, e.g., 90\% confidence intervals. One can show that the minimum is achieved for the values of $\theta$ closest to $\pi/2$ and for at least one of the four combinations of $\varphi_{1,3}=\varphi_{1,3}^{\min(\max)}$ and $\varphi_{2,4}=\varphi_{2,4}^{\max(\min)}$ since they minimize the last $\sin^2\theta$--term which is dominant for $\theta\sim\pi/2$ and $\varphi_{2(4)}-\varphi_{1(3)}\sim\pi$. This yields $F_{\min}=0.9975$ and $\Delta_{\max}=(1-\sqrt{F_{\min}})/2=6\times10^{-4}$. In this case the critical transmission distance is limited to 40\,km and the maximum $\ell_\sec/\ell_\ver$ ratio is about 0.4 at zero loss. A similar estimation of $F$ and $\Delta$ using the upper-bounded phase modulation errors is made in the GLLP security analysis for the phase-encoding decoy-state BB84 with source flaws~\cite{Xu15}. For completeness, on the right-hand side of Fig.~\ref{fig:r_sec_PM} we plot the $\ell_\sec/\ell_\ver$ ratio for various $\Delta$--values, from which one can see that $\Delta\lesssim10^{-8}$ is required in order to make the basis-dependence effect negligible.

\section{Conclusions}\label{sec:conclusions}

In this report, we study the security of practical efficient decoy-state BB84 QKD protocol with polarization encoding accounting for the source flaws due to imperfect intensity and phase modulation. Based on our experimental data, we use the normal distribution to describe the non-Poissonian photon-number statistics and derive the generalized bounds on zero/single-photon yields. We find that the effect of intensity fluctuations on the secret key length is less than 5\% for any transmission distances up to 100\,km.
It is also demonstrated that for our experimental setup the proposed method turns out to be less conservative than the one introduced in Refs.~\cite{Wang08,Wang09}.
In order to take into account the imperfect polarization state preparation, we apply an idea from the fully-passive source approach \cite{Zapatero23} and compute the density matrices of mixed states in two polarization bases; then we evaluate the fidelity between them and estimate the upper bound on the phase error rate using the concept of imbalanced quantum coin \cite{GLLP,Koashi09,Lo07}. From the analysis of experimental angular distributions we find that the key length is reduced by less than 9 and 47\% for the transmission distances up to 50 and 100\,km, respectively. If the phase modulation uncertainties are taken into account, the ultimate key transmission distance reduces from 160\,km to 120\,km.
The method of estimating the fidelity between the mixed states turns out to be significantly better than the more conservative evaluation from minimizing the fidelity of pure states (e.g., the maximum distance of 120\,km versus 40\,km).
We emphasize that these results are preliminary. The fidelity estimation could be further improved in our future work by (i) enlarging the amount of data for the statistical analysis (we have just $\sim$2500 data points for each polarization state), (ii)~improving the voltage and temperature control, (iii) applying a more advanced post-selection procedure for the polarimeter data, (iv) performing a more sophisticated angular analysis.
Although the obtained numerical results are valid only for the particular simple optical layout that mimics a realistic QKD transmitter, the explicit analytical formulas of the present approach can be readily applied to any practical implementation of the widely-used two-decoy-state BB84 QKD protocol with imperfect quantum state preparation augmented with the relevant source characterization data.

\section{Acknowledgments}
This work is supported by the Priority 2030 program at the National University of Science and Technology ``MISIS'' under the project K1-2022-027.

\begin{appendices}

\section{Generalized bounds on $\bm{\Y_{0,1}}$}\label{app:new_Y}

Following the procedure for classical Poissonian source in Ref.~\cite{Ma05}, one can write
\begin{equation}
    \begin{split}
        Q_{\nu_2} P_{1|\nu_1} - Q_{\nu_1} P_{1|\nu_2} &= (P_{0|\nu_2} P_{1|\nu_1} - P_{0|\nu_1} P_{1|\nu_2}) \Y_0 \\
        &+ \sum_{n=2}^{\infty} (P_{n|\nu_2} P_{1|\nu_1} - P_{n|\nu_1} P_{1|\nu_2}) \Y_n \\
        &\leq (P_{0|\nu_2} P_{1|\nu_1} - P_{0|\nu_1} P_{1|\nu_2}) \Y_0 \,,
    \end{split}
    \label{eq:Y0_derivation}
\end{equation}
where the inequality follows from the required condition
\begin{equation}
    P_{n|\nu_2} P_{1|\nu_1} - P_{n|\nu_1} P_{1|\nu_2} = \iint_0^\infty e^{-\nu_1 - \nu_2} \frac{\nu_1 \nu_2}{n!}
    (\nu_2^{n-1} - \nu_1^{n-1}) G_{\nu_1} G_{\nu_2} d\nu_1 d\nu_2 \stackrel{n\geq2}{<} 0 \,.
    \label{eq:check1}
\end{equation}
Here, for short, we use the notation $G_\alpha\equiv G(\alpha,\alphabar,\sigma_\alpha)/\int_0^\infty G(\alpha,\alphabar,\sigma_\alpha)d\alpha$. Thus, the modified lower bound on $\Y_0$ is given by
\begin{equation}
    \Y_0 \geq \Y_0^l = \max\bigg\{ \frac{Q_{\nu_2} P_{1|\nu_1} - Q_{\nu_1} P_{1|\nu_2}}{P_{0|\nu_2} P_{1|\nu_1} - P_{0|\nu_1} P_{1|\nu_2}} ,\, 0 \bigg\} \,.
    \label{eq:Y0_l_new}
\end{equation}

Writing another combination of gains, one obtains
\begin{equation}
    \begin{split}
        Q_{\nu_1} P_{0|\nu_2} - Q_{\nu_2} P_{0|\nu_1} &= (P_{1|\nu_1} P_{0|\nu_2} - P_{1|\nu_2} P_{0|\nu_1}) \Y_1 \\
        &+ \sum_{n=2}^{\infty} (P_{n|\nu_1} P_{0|\nu_2} - P_{n|\nu_2} P_{0|\nu_1}) \Y_n \\
        &\leq (P_{1|\nu_1} P_{0|\nu_2} - P_{1|\nu_2} P_{0|\nu_1}) \Y_1 \\
        &+ \frac{P_{2|\nu_1} P_{0|\nu_2} - P_{2|\nu_2} P_{0|\nu_1}}{P_{2|\mu}} \sum_{n=2}^{\infty} P_{n|\mu} \Y_n \\
        &= (P_{1|\nu_1} P_{0|\nu_2} - P_{1|\nu_2} P_{0|\nu_1}) \Y_1 \\
        &+ \frac{P_{2|\nu_1} P_{0|\nu_2} - P_{2|\nu_2} P_{0|\nu_1}}{P_{2|\mu}} (Q_{\mu} - P_{0|\mu} \Y_0 - P_{1|\mu} \Y_1) \,,
    \end{split}
\end{equation}
where the inequality follows from another required condition,
\begin{equation}
    \begin{split}
        & P_{n|\mu}(P_{2|\nu_1} P_{0|\nu_2} - P_{2|\nu_2} P_{0|\nu_1}) - P_{2|\mu}(P_{n|\nu_1} P_{0|\nu_2} - P_{n|\nu_2} P_{0|\nu_1}) \\
        &= \iiint_0^\infty e^{-\mu -\nu_1 - \nu_2} \frac{\mu^{n+2}}{2! n!}
        \bigg[\frac{\nu_1^2}{\mu^2} - \frac{\nu_2^2}{\mu^2} - \bigg(\frac{\nu_1^n}{\mu^n} - \frac{\nu_2^n}{\mu^n}\bigg)\bigg]
        G_\mu G_{\nu_1} G_{\nu_2} d\mu d\nu_1 d\nu_2 \stackrel{n\geq2}{\geq} 0 \,.
    \end{split}
    \label{eq:check2}
\end{equation}
As a result, the lower bound on the single-photon yield is given by
\begin{equation}
    \Y_1 \geq \Y_1^l =  \frac{Q_{\nu_1} P_{0|\nu_2} - Q_{\nu_2} P_{0|\nu_1} - \displaystyle\frac{P_{2|\nu_1} P_{0|\nu_2} - P_{2|\nu_2} P_{0|\nu_1}}{P_{2|\mu}} (Q_{\mu} - P_{0|\mu} \Y_0^l)}{P_{1|\nu_1} P_{0|\nu_2} - P_{1|\nu_2} P_{0|\nu_1} - \displaystyle\frac{P_{2|\nu_1} P_{0|\nu_2} - P_{2|\nu_2} P_{0|\nu_1}}{P_{2|\mu}} P_{1|\mu}} \,.
    \label{eq:Y1_l_new}
\end{equation}
Substituting $P_{n|\alpha}\to e^{-\alphabar}\,\alphabar^n/n!$ in Eqs.~\eqref{eq:Y0_l_new},\eqref{eq:Y1_l_new}, one easily recovers the ``classic'' yield expressions for a perfect Poissonian source \cite{Ma05},
\begin{equation}
	\Y_0^l = \max \bigg\{ \frac{\overline\nu_1 Q_{\nu_2} e^{\overline\nu_2} - \overline\nu_2 Q_{\nu_1} e^{\overline\nu_1}}{\overline\nu_1 - \overline\nu_2} \,, 0 \bigg\} \,,
	\label{eq:Y0_l_Poisson}
\end{equation}
\begin{equation}
	\Y_1^l = \frac{\overline\mu}{(\overline\nu_1 - \overline\nu_2) (\overline\mu - \overline\nu_1 - \overline\nu_2)} \bigg[ Q_{\nu_1} e^{\overline\nu_1} - Q_{\nu_2} e^{\overline\nu_2} - \frac{\overline\nu_1^2 - \overline\nu_2^2}{\overline\mu^2} \big( Q_\mu e^{\overline\mu} - \Y_0^l \big) \bigg] \,.
	\label{eq:Y1_l_Poisson}
\end{equation}
Also, imposing $P_{n|\nu_2}=\delta_{n0}$ (i.e. $\rho_{\nu_2}=\ketbra{0}{0}$) and $P_{2|\nu_1}/P_{2|\mu}\geq P_{n|\nu_1}/P_{n|\mu}$ for $n\geq2$, we recover the result of Ref.~\cite{Foletto22}.

Note that $\{\mu,\nu_1,\nu_2\}$ are random variables, and, hence, one cannot justify the inequalities \eqref{eq:check1} and \eqref{eq:check2} analytically by simply imposing the constraints $\nu_2<\nu_1$ and $\nu_1+\nu_2<\mu$, as was done in Ref.~\cite{Ma05}. Therefore, we verify them numerically by computing the integrals in terms of special functions. For completeness, we present $P_{n|\alpha}$ \eqref{eq:P_n} for $n=0,1,2$ expressed in terms of the error function :

\begin{equation}
    \begin{split}
        P_{0|\alpha} &= e^{-\alphabar + \frac{\sigma_\alpha^2}{2}} \bigg[1 - \erf\bigg(-\frac{\alphabar - \sigma_\alpha^2}{\sqrt2 \sigma_\alpha}\bigg)\bigg] \bigg[1 + \erf\bigg(\frac{\alphabar}{\sqrt{2}\sigma_\alpha}\bigg)\bigg]^{-1} \,, \\
        P_{1|\alpha} &= \frac{e^{-\frac{\alphabar^2}{2\sigma_\alpha^2}}}{\sqrt{2\pi}}
        \bigg\{2\sigma_\alpha +
        \sqrt{2\pi} e^{\frac{(\alphabar - \sigma_\alpha^2)^2}{2\sigma_\alpha^2}} (\alphabar - \sigma_\alpha^2) \bigg[1 + \erf\bigg(\frac{\alphabar - \sigma_\alpha^2}{\sqrt2 \sigma_\alpha}\bigg)\bigg]\bigg\} \\
        & \times \bigg[1 + \erf\bigg(\frac{\alphabar}{\sqrt{2}\sigma_\alpha}\bigg)\bigg]^{-1} \,, \\
        P_{2|\alpha} &= \frac{e^{-\frac{\alphabar^2}{2\sigma_\alpha^2}}}{2\sqrt{2\pi}} 
        \bigg\{2\sigma_\alpha (\alphabar - \sigma_\alpha^2) +
        \sqrt{2\pi} e^{\frac{(\alphabar - \sigma_\alpha^2)^2}{2\sigma_\alpha^2}} \big(\sigma_\alpha^2 + (\alphabar - \sigma_\alpha^2)^2\big) \\
        & \times\bigg[1 + \erf\bigg(\frac{\alphabar - \sigma_\alpha^2}{\sqrt2 \sigma_\alpha}\bigg)\bigg]\bigg\}
        \bigg[1 + \erf\bigg(\frac{\alphabar}{\sqrt{2}\sigma_\alpha}\bigg)\bigg]^{-1} \,.
    \end{split}
\end{equation}
Using Wolfram Mathematica 12 (with numerical verification), $P_{n|\alpha}$ with arbitrary $n$ can be written in terms of the gamma function $\Gamma$ and the confluent hypergeometric function of the first kind of ${_1F_1}$,

\begin{equation}
    \begin{split}
        P_{n|\alpha} &= \frac{2^{\frac{n+1}{2}} e^{-\frac{\alphabar^2}{2\sigma_\alpha^2}} \sigma_\alpha^{n-1}}{\sqrt{2\pi}\, n!} \bigg\{ \sigma_\alpha \Gamma\bigg(\frac{n+1}{2}\bigg) {_1F_1} \bigg(\frac{n+1}{2}, \frac{1}{2}, \frac{(\alphabar - \sigma_\alpha^2)^2}{2\sigma_\alpha^2}\bigg) \\
        &+ \sqrt{2} (\alphabar - \sigma_\alpha^2) \Gamma\bigg(\frac{n+2}{2}\bigg) {_1F_1} \bigg(\frac{n+2}{2}, \frac{3}{2}, \frac{(\alphabar - \sigma_\alpha^2)^2}{2\sigma_\alpha^2}\bigg) \bigg\} \\
        &\times \bigg[1 + \erf\bigg(\frac{\alphabar}{\sqrt{2}\sigma_\alpha}\bigg)\bigg]^{-1} \,.
    \end{split}
    \label{eq:P_nalpha}
\end{equation}

\section{Secret key length}\label{app:l_sec}

We consider the efficient BB84 protocol \cite{Lo05a} with the basis choice probabilities of $p_X=0.9$ and $p_Y=0.1$, in which the $X$--basis is used for the secret key distillation while the $Y$--basis is used for the phase error estimation. Since the simultaneous rotation of the bases around the $z$--axis does not affect the security, here for simplicity we use the notation $\{X,Y\}$ instead of physical $\{\Xp,\Yp\}$. The length of the secret key is estimated as follows~\cite{Trushechkin17,Zhang17},
\begin{equation}
	\ell_\sec^X = m_1^{X,l} \big[1 - h_2\big(E_1^{\ph,X,u}\big)\big] - \text{leak} - 5\log_2\frac{1}{\varepsilon_\pa} \,,
	\label{eq:l_sec}
\end{equation}
where the first and the last terms represent the privacy amplification step and are determined by $m_1^{X,l}$ -- the lower bound on the number of bits in the verified key of length $\ell_\ver^X$ obtained from signal single-photon pulses, $E_1^{\ph,X,u}$ -- the upper bound on the single-photon phase error rate in the $X$--basis, and $\varepsilon_\pa=10^{-12}$ -- the tolerable failure probability for the privacy amplification step. The $h_2$--function is the standard Shannon binary entropy. The second term in \eqref{eq:l_sec} is the amount of information about the sifted key leaked to Eve during the error correction and verification steps, that can be parametrized as
\begin{equation}
	\text{leak} = \ell_\sift^X f_\ec(\hat{E}_\mu^X) h_2(\hat{E}_\mu^X) + \ell_\text{hash} \,,
	\label{eq:leak}
\end{equation}
where $\ell_\sift^X$ is the sifted key length, $\hat{E}_\mu^X$ is the signal quantum bit error rate (QBER), determined during the error correction step, $f_\ec\geq1$ is the error correction code inefficiency, and $\ell_\text{hash}$ is the verification hash-tag length. Usually in the literature $f_\ec\sim1.2$ is assumed, however, for realistic codes it can be significantly higher for low QBER. Therefore, we extract the $f_\ec(E_\mu)$--dependence of the low-density parity-check (LDPC) codes from Ref.~\cite{Borisov22} (see the upper-left SEC plot in Fig.~3 in \cite{Borisov22}).

The finite-key-size effects and statistical fluctuations are taken into account in our analysis. According to the central limit theorem, the binomial distributions of $M_\alpha^X\sim\text{Bi}(N_\alpha^X,Q_\alpha^X)$ and $m_1^X\sim\text{Bi}(\ell_\ver^X,Q_1^X/Q_\mu^X)$ can be well approximated by the normal distribution. The upper and lower bounds on $Q_\alpha^X$ and $m_1^X$ are evaluated using the Wald confidence interval~\cite{Trushechkin17},
\begin{equation}
	Q_\alpha^{X,u(l)} = \hat{Q}_\alpha^X \pm z_{1-\varepsilon} \sqrt\frac{\hat{Q}_\alpha^X (1 - \hat{Q}_\alpha^X)}{N_\alpha^X} \,, \quad
    \hat{Q}_\alpha^X = \frac{M_\alpha^X}{N_\alpha^X} \,,
	\label{eq:Q_ul}
\end{equation}
\begin{equation}
	m_1^{X,l} = \ell_\ver^X \frac{Q_1^{X,l}}{Q_\mu^{X,u}}  - z_{1-\varepsilon} \sqrt{\ell_\ver^X \frac{Q_1^{X,l}}{Q_\mu^{X,u}} \bigg( 1 - \frac{Q_1^{X,l}}{Q_\mu^{X,u}} \bigg)} \,, \quad
    Q_1^{X,l} = P_{1|\mu} \Y_1^{X,l} \,,
	\label{eq:m1}
\end{equation}
where $N_\alpha^X$ and $M_\alpha^X$ are the total counts (before sifting) of transmitted and detected pulses of given intensity, respectively.

It is well known that for the basis-independent source ($\rho_X=\rho_Y$) in the asymptotic limit ($N_\alpha\to\infty$) the phase error rate in the $X$--basis is equal to the bit error rate in the $Y$--basis. If $\rho_X\neq\rho_Y$ and the data samples are finite, the phase error bound receives two corrections,
\begin{equation}
	E_1^{\ph,X,u} = E_1^{Y,u} + \Theta_\text{coin} + \Theta_\text{stat} = \widetilde{E}_1^{Y,u} + \Theta_\text{stat} \,,
    \label{eq:E1ph_finite-key}
\end{equation}
where the basis-dependence correction $\Theta_\text{coin}$ is parametrized in terms of the effective quantum coin imbalance $\Delta^\prime$ \cite{Lo07},
\begin{equation}
    \Theta_\text{coin} = 4 \Delta^\prime (1 - \Delta^\prime) (1 - 2E_1^{Y,u}) + 4(1 - 2\Delta^\prime) \sqrt{\Delta^\prime (1 - \Delta^\prime) E_1^{Y,u} (1 - E_1^{Y,u})} \,,
    \label{eq:Theta_coin}
\end{equation}
\begin{equation}
    \Delta^\prime = \frac{1 - \sqrt{F(\rho_X,\rho_Y)}}{\Y_1^{X,l} + \Y_1^{Y,l}} \,,
\end{equation}
and $\Theta_\text{stat}$ takes into account different sizes of two random data samples and statistical fluctuations and is determined by numerical solution of the following equation~\cite{Zhang17},
\begin{equation}
	\sqrt{\frac{m_1^{Y,l} + m_1^{X,l}}{\widetilde{E}_1^{Y,u} (1 - \widetilde{E}_1^{Y,u}) m_1^{Y,l} m_1^{X,l}}}
	~ 2^{-(m_1^{Y,l} + m_1^{X,l}) \xi(\Theta_\text{stat})} = \varepsilon \,,
    \label{eq:Theta_stat1}
\end{equation}
\begin{equation}
	\begin{split}
		\xi(\Theta_\text{stat}) &= h_2\bigg(\widetilde{E}_1^{Y,u} + \frac{m_1^{X,l}}{m_1^{Y,l} + m_1^{X,l}} \Theta_\text{stat} \bigg)
		-\frac{m_1^{Y,l}}{m_1^{Y,l} + m_1^{X,l}} h_2(\widetilde{E}_1^{Y,u}) \\
        &- \frac{m_1^{X,l}}{m_1^{Y,l} + m_1^{X,l}} h_2(\widetilde{E}_1^{Y,u} + \Theta_\text{stat}) \,.
	\end{split}
	\label{eq:Theta_stat2}
\end{equation} 

As pointed out in Ref.~\cite{Trushechkin17}, in general, one cannot use binomial distribution for QBER since if Eve performs a coherent attack, the errors in different positions in the key cannot be treated as independent events. Therefore, the upper bound on the single-photon bit error rate is estimated differently compared to Ref.~\cite{Zhang17},
\begin{equation}
	E_1^{Y,u} = \frac{\ell_\sift^Y \hat{E}_\mu^Y - \overline{m}_0^{Y,l}}{m_1^{Y,l}} \,,
	\label{eq:E1_u}
\end{equation}
where the number of bit errors, obtained from the zero-photon pulses due to background events, is distributed as $\overline{m}_0^Y\sim\text{Bi}(N_\mu^Y,p_Y P_{0|\mu}\Y_0^Y/2)$ and lower bounded by
\begin{equation}
	\overline{m}_0^{Y,l} = N_\mu^Y p_Y \frac{P_{0|\mu} \Y_0^{Y,l}}{2} - z_{1-\varepsilon} \sqrt{N_\mu^Y p_Y \frac{P_{0|\mu} \Y_0^{Y,l}}{2} \bigg( 1 - p_Y \frac{P_{0|\mu} \Y_0^{Y,l}}{2} \bigg)} \,.
	\label{eq:m0}
\end{equation}
Since the $Y$--basis is used only for parameter estimation, the signal QBER $\hat{E}_\mu^Y$ is computed directly from public disclosure of the sifted keys generated in the $Y$--basis. All other quantities in the $Y$--basis are computed the same way as in the $X$--basis with replacement $\ell_\ver^X\to\ell_\sift^Y$ in Eq.~\eqref{eq:m1}.

One can count 14 statistical confidence bounds in total that are  required to compute $\ell_\sec^X$. Therefore, in order to have the estimation \eqref{eq:l_sec} satisfied with the probability not less than $1-\varepsilon_\text{decoy}$, one has to set the failure probability of each bound  $\varepsilon=\varepsilon_\text{decoy}/14$. We set $\varepsilon_\text{decoy}=10^{-12}$. For completeness, taking into account also the verification failure probability $\varepsilon_\ver$ (e.g., using the $\epsilon$--universal polynomial hashing, $\varepsilon_\ver\leq2\times10^{-11}$ for $\ell_\sift\simeq10^6$ and $\ell_\text{hash}=50$ \cite{Fedorov18}), the overall (in)security parameter of the entire QKD system is equal to $\varepsilon_\text{tot}=\varepsilon_\text{decoy}+\varepsilon_\ver+\varepsilon_\pa$.

For the numerical simulation of $\hat{Q}_\alpha$ and $\hat{E}_\alpha$, we use the simplest theoretical model from Ref.~\cite{Ma05},
\begin{equation}
    Q_\alpha = 2p_\dc + 1 - e^{-t\eta\alpha} \,, \quad
    E_\alpha = \frac{p_\dc + p_\opt (1 - e^{-t\eta\alpha})}{2p_\dc + 1 - e^{-t\eta\alpha}} \,,
\end{equation}
with the single-photon detector quantum efficiency of $\eta=10\%$ and dark count probability of $p_\dc=10^{-6}$, photon transmission probability of $t=10^{-(\beta L+3)/10}$ determined by Bob's internal loss of 3\,dB and quantum channel loss coefficient $\beta\simeq0.2$\,dB/km, and optical misalignment error of $p_\opt=1\%$. For the statistical analysis, we set
\begin{equation}
	N_\mu^X = \frac{\ell_\sift^X}{p_X Q_\mu} \,, \quad
	N_{\nu_i}^X = N_\mu^X \, \frac{p_{\nu_i}}{p_\mu} \,, \quad
    N_\alpha^Y = N_\alpha^X \frac{p_Y}{p_X} \,, \quad
    \ell_\sift^Y = \ell_\sift^X \frac{p_Y^2}{p_X^2} \,,
	\label{eq:N_th}
\end{equation}
where we take the $X$--basis sifted key block length of $\ell_\sift^X=1.36\times10^6$ \cite{Borisov22}. We also neglect the error correction and verification failure probabilities and set $\ell_\ver^X=\ell_\sift^X$.

\end{appendices}

\bibliographystyle{utphys}
\bibliography{bibliography}

\end{document}